\documentclass[runningheads]{llncs}
\usepackage{graphicx}
\usepackage{xcolor}
\usepackage{amsmath}
\usepackage{amssymb}
\usepackage{multirow}
\usepackage{booktabs}
\usepackage{footnote}
\makesavenoteenv{tabular}
\makesavenoteenv{table}
\usepackage{cite}
\usepackage[ruled,vlined]{algorithm2e}
\usepackage{float}
\usepackage{hyperref}

\makeatletter
\newcommand{\printfnsymbol}[1]{%
  \textsuperscript{\@fnsymbol{#1}}%
}
\makeatother

\title{Diversity Aware Relevance Learning for Argument Search}
 \author{
Michael Fromm\thanks{equal contribution}\inst{1} 
\and
Max Berrendorf\printfnsymbol{1} \inst{1} 
\and
Sandra Obermeier \inst{1} 
\and
Thomas Seidl \inst{1} 
\and
Evgeniy Faerman \inst{1}
}

 \authorrunning{Fromm et al.}

 \institute{Database Systems and Data Mining, LMU Munich, Germany  \\
 \email{fromm@dbs.ifi.lmu.de}}

\usepackage[acronym]{glossaries}

\newacronym{argument-mining}{AM}{Argument Mining}
\newacronym{arg_ret_sys}{ARS}{Argument Retrieval System}
\newacronym{bert-based-premise-representation}{BERT}{BERT}
\newacronym{claim-based-premise-representation}{CLAIM-SIM}{CLAIM-SIM}
\newacronym{relevance-model}{relevance-model}{relevance model}

\newacronym{dumani-first512}{first512}{Dumani first512}
\newacronym{dumani-sentences}{sentences}{Dumani sentences}
\newacronym{dumani-sliding-window}{sliding}{Dumani sliding}
\newacronym{bert-zero-shot-knn}{BERT Zero-Shot}{BERT Zero-Shot}
\newacronym{learned-similarity-knn}{Learned Similarity}{Learned Similarity}
\newacronym{biased-coreset}{Biased Coreset}{Biased Coreset}

\newacronym{bert-zero-shot-clustered}{BERT Zero-Shot + Cluster}{}

\DeclareMathOperator*{\argmax}{argmax}

\glsdisablehyper

\begin{document}
\maketitle              

\begin{abstract}
In this work, we focus on retrieving relevant arguments for a query claim covering diverse aspects.
State-of-the-art methods rely on explicit mappings between claims and premises and thus cannot utilize extensive available collections of premises without laborious and costly manual annotation.
Their diversity approach relies on removing duplicates via clustering, which does not directly ensure that the selected premises cover all aspects.
This work introduces a new multi-step approach for the argument retrieval problem. Rather than relying on ground-truth assignments, our approach employs a machine learning model to capture semantic relationships between arguments.
Beyond that, it aims to cover diverse facets of the query instead of explicitly identifying duplicates. 
Our empirical evaluation demonstrates that our approach leads to a significant improvement in the argument retrieval task, even though it requires fewer data than prior methods.
Our code is available at \url{https://github.com/fromm-m/ecir2021-am-search}.
\end{abstract}


\section{Introduction}
Argumentation is a paramount process in society, and debating on socially relevant topics requires high-quality and relevant arguments.
In this work, we deal with the problem of \emph{argument search}, which is also known as \emph{argument retrieval}.
The goal is to develop an \acrfull{arg_ret_sys} which organizes arguments, previously extracted from various sources  \cite{chernodub2019targer,stab-etal-2018-cross,trautmann2020relational, ein2020corpus}, in an accessible form.
Users then formulate a query to access relevant arguments retrieved by the \acrshort{arg_ret_sys}. 
The query can be defined as a \emph{topic}, e.g. \textit{Energy} in which case the \acrshort{arg_ret_sys} retrieves all possible arguments without further specification~\cite{trautmann2020relational, stab-etal-2018-cross, DBLP:conf/webi/FrommF019}.
Our work deals with a more advanced case, where a query is formulated in the form of a \emph{claim}, and the user expects \emph{premises} attacking or supporting this query claim. 
An example of a claim related to the topic \textit{Energy} could be \textit{``We should abandon Nuclear Energy"} and a supporting premise, e.g., \textit{``Accidents caused by Nuclear Energy have longstanding negative impacts"}.
A popular search methodology to find relevant premises is a similarity search, where the representations of the retrieved premises are similar to the representation of the (augmented) query claim~\cite{akiki:2020, feger2020structure, Staudte2020SentArgAH, wachsmuth2020overview}.
However, as noted by~\cite{dumani2019systematic,dumani2020framework}, the relevance of a premise does not necessarily coincide with pure text similarity. 
Therefore, the authors of \cite{dumani2020framework} advocate to utilize the similarity between the query claim and other claims in an \acrshort{arg_ret_sys} database and retrieve the premises assigned to the most similar claims.
However, such \acrshort{arg_ret_sys} requires ground truth information about the premise to claim assignments and therefore has limited applicability:
Either the information sources are restricted to those sources where such information is already available or can automatically be inferred, or expensive human annotations are required.
To mitigate this problem and keep the original system's advantages, we propose to use a machine learning model \emph{to learn} the relevance between premises and claims.
Using this model, we can omit the (noisy) claim-claim matching step and evaluate the importance of (preselected) candidate premises directly for the query claim.
Since the relevance is defined on the semantic level, we have to design an appropriate training task to enable the model to learn semantic differences between relevant and non-relevant premises.
Furthermore, an essential subtask for an \acrshort{arg_ret_sys} is to ensure that the retrieved premises do not repeat the same ideas. 
Previous approaches~\cite{dumani2020framework} employ clustering to eliminate duplicates. 
However, clustering approaches often group data instances by other criteria than expected by the users~\cite{kriegel2009clustering}, as also observed in \gls{argument-mining} applications~\cite{reimers-etal-2019-classification}. 
For our method, we propose an alternative to clustering based on the idea of \emph{core-sets}~\cite{sener2017active}, where the goal is to cover the space of relevant premises as well as possible.

\section{Preliminaries}
In our setting, the query comes in the form of a claim, and an answer is a sorted list of \emph{relevant} premises from the \acrshort{arg_ret_sys} database.
A premise is considered relevant if it attacks or supports the idea expressed in the claim~\cite{wachsmuth2017computational,habernal-gurevych-2016-argument}.
We denote the query claim by $c_{query}$ and the list of premises retrieved by \acrshort{arg_ret_sys} by $A$, with the length being fixed to $|A| = k$.
Besides relevance, another vital requirement for the \acrshort{arg_ret_sys} is that premises in $A$ should have diverse semantic meaning.
We consider a two-step retrieval process.
First, in the \emph{pre-filtering}, the system selects a set of candidate premises $\mathcal{T}$ with $|\mathcal{T}| > k$.
This step should have a relatively high recall, i.e., find most of the relevant premises.
For a fair comparison to previous approaches, we leave the pre-filtering step from~\cite{dumani2020framework} unchanged.
We note that the current version of pre-filtering requires ground-truth matchings of premises to claims restricting its applicability and improving it in future work.
The pre-filtering process described in~\cite{dumani2020framework} has several steps.
When a query claim arrives, the system first determines \emph{claims} from the database which have the highest Divergence from Randomness~\cite{amati2002probabilistic} similarity to the query claim. 
Next, the system receives the corresponding claim clusters of the claims found in the previous step, and all premises assigned to all claims from these clusters are collected in a candidate seed set $\mathcal{T}_{seed}$.
Each premise $p \in \mathcal{T}_{seed}$ is then used as a query to obtain the most similar premises using the BM25 score, which are accumulated in a set $\mathcal{T}_{sim}$.
The complete candidate set is then given as the union $\mathcal{T} = \mathcal{T}_{seed} \cup \mathcal{T}_{sim}$.

\section{Our Approach for Candidate Refinement}
Our work's primary focus is the second step in the retrieval process or the candidate refinement/ranking procedure. 
The candidates are analyzed more thoroughly in the refinement step, and non-relevant or redundant premises are discarded.
Our refinement process comprises two components.
The \emph{relevance filter} component determines each premise's relevance from the candidate set $\mathcal{T}$ using an advanced machine learning model that keeps only the most relevant ones. 
The relevance filter thus maps the candidate set $\mathcal{T}$ to a subset thereof, denoted by $\mathcal{T}_{filtered} \subseteq \mathcal{T}$.
The subsequent \emph{premise ranker} selects and orders $k$ premises from $\mathcal{T}_{filtered}$ to the result list $A$.
An essential requirement for the premise ranker is that $A$ does not contain semantically redundant premises. 
In the following, we describe both components in more detail.

\subsection{Relevance Filter} \label{relevance_filter}
\paragraph{Inference} 
Given a set of candidate premises $\mathcal{T}$ and the query claim $c_{query}$, the \emph{relevance filter} determines the relevance score of each candidate $p \in \mathcal{T}$ denoted as $r(p \mid c_{query})$.
We keep only the most relevant candidates in the filtered candidate set
$
\mathcal{T}_{filtered} = \{ p \in \mathcal{T} \mid r(p \mid c_{query}) > \tau \}
$
with a relevance threshold $\tau$.
We interpret the relevance prediction as a binary classification problem and train a Transformer~\cite{vaswani2017attention} model to solve this classification task given the concatenation of the candidate premise and the query claim.
At inference time, we use the predicted likelihood as the relevance score and evaluate the model on the concatenation of each candidate premise with the query claim.

\paragraph{Training Task}
For the training part, we assume that we have access to a (separate) dataset $D=(\mathcal{P}', \mathcal{C}', \mathcal{R}^+)$ containing a set of premises $\mathcal{P}'$, a set of claims $\mathcal{C}'$ and a set of relevant premise-claim pairs $\mathcal{R}^+ \subseteq \mathcal{P}' \times \mathcal{C}'$.
In fact, several datasets fulfill this requirement, e.g.,~\cite{wachsmuth2017building, dumani2019systematic}.
Since the relevance filter receives as input the remaining candidate premises after the pre-filtering, we assume that the non-relevant premises appear similar to the relevant ones.
Therefore, the training task must be designed very carefully to enable the model to learn semantic differences between relevant and non-relevant premises.
We use the ground truth premise-claim pairs $\mathcal{R}^+$ as instances of the positive class (i.e., an instance of matching pairs). 
For each positive instance $(p^+, c) \in \mathcal{R}^+$, we generate $L$ instances of the negative class $(p_i^-, c) \in \mathcal{R}^-$.
For $p_i^-$, we choose the $L$ most similar premises according to a premise similarity $psim$, which do not co-occur with $c$ in the database.
We use the cosine similarity $psim(p, p') = \cos (\phi(p), \phi(p'))$ between the premise representations $\phi(p)$ obtained from a pre-trained BERT model without any fine-tuning as premise similarity.\footnote{Using average pooling of the second-to-last hidden layer over all tokens}
The transformer model, which predicts the premise-claim relevance, is initialized with weights from a pre-trained BERT model~\cite{devlin-etal-2019-bert}.

\subsection{Premise Ranker}
\begin{algorithm}[t]
\caption{Biased Coreset}
\label{alg:biased_coreset}
\SetAlgoLined
\KwData{candidates $\mathcal{T}$, relevances $R$, similarity $psim$, $k \in \mathbb{N}$, $\alpha \in [0, 1]$}
\KwResult{premise list $A$}
\For{$i=1$ \KwTo $k$}{
    \lIf{$|A| = 0$}{$a =\argmax \limits_{p \in \mathcal{T}} \alpha \cdot R[p]$}
    \lElse{$a = \argmax \limits_{p \in \mathcal{T}} \alpha \cdot R[p] - (1 - \alpha) \cdot \max \limits_{a \in A} psim(a, p)$}
    $A.append(a)$; $\mathcal{T} = \mathcal{T} \setminus \{a\}$
}
\end{algorithm}
The \emph{premise ranker} receives a set of relevant premises with the corresponding relevance scores and makes the final decision about the premises and the order they are returned to the user. 
Since the two relevance filtering steps have been applied, we assume that most remaining candidates are relevant. 
Thus, the main task of this component is to avoid semantic duplicates. 
While related approaches~\cite{dumani2020framework} advocate for the utilization of clustering for the detection of duplicates and expect that premises with the same meaning end up in the same clusters, we pursue a different idea.
Instead of explicitly detecting the duplicates, we aim to identify $k$ premises that adequately represent all premises in $T_{filtered}$.
Therefore, we borrow the idea of core-sets from \cite{sener2017active} and aim to select $k$ premises from the final candidate set $\mathcal{T}_{filtered}$ such that for each candidate premise $p \in \mathcal{T}_{filtered}$ there is a similar premise in the result $A$.
More formally, we denote
$
    Q(p, A) = \max_{a \in A} psim(p, a)
$
as a measure of how well $p$ is represented by $A$, using the premise similarity $psim$.
Thus,
$
    \bar{Q}(A) = \min_{p \in \mathcal{T}_{filtered}} Q(p, A)
$
denotes the worst representation of any premise $p \in \mathcal{T}_{filtered}$ by $A$.
Hence, we aim to maximize $\bar{Q}$ such that every premise $p$ is well represented.
This min-max objective ensures that every premise is well-represented at not only the majority of premises.
To solve the selection problem, we adopt the greedy approach from~\cite{sener2017active}. 
Since our goal is not only that the selected premises represent the remaining candidates well, but also that the selected premises have high relevance, we start with the most relevant premise and also consider the relevance score $r$ for the next assignments, with a weighting parameter $\alpha \in [0, 1]$.
$\alpha = 0$ scores only according to the coreset criterion, while $\alpha = 1$ uses only the relevance.
The full algorithm is presented in Algorithm~\ref{alg:biased_coreset}.

\paragraph{Premise Representation}
The premise ranker requires a meaningful similarity measure to compare premises with each other. 
As also noted in~\cite{dumani2020framework}, semantically similar premises might often be expressed differently.
Therefore, an essential requirement for the similarity function is that it captures semantic similarities. 
We investigate two approaches to obtain vector representations on which we compute similarities using l1, l2, or cos similarity. 
Previous works demonstrated that BERT models pre-trained on language modeling can capture argumentative context~\cite{DBLP:conf/webi/FrommF019}.
Thus, our first \emph{\acrshort{bert-based-premise-representation}} similarity function employs a BERT model without fine-tuning to encode the premises.
We abbreviate these representations with \emph{\acrshort{bert-based-premise-representation}}.
As an alternative, we propose representing each premise by a vector of relevance scores to selected claims in the database.
While we can use randomly selected claims or cluster all claims in the database, many databases already contain topic information about the claims, such as e.g., "Energy."
Thus, we restrict the selection of claims for each premise to the same high-level topic of interest. In this case, all premises retrieved for a single query belong to the same topic.  We do not consider it a substantial restriction since arguments always exist in some context, and it rarely makes sense to retrieve premises from different topics for the same query. 
We utilize our relevance filter model to compute relevance scores for the premise and each of the selected claims.
We call the resulting vector of stacked similarities \emph{\acrshort{claim-based-premise-representation}} representation. 
We hypothesize that a similar relationship to the selected claims is a good indicator of semantically similar premises.
\section{Evaluation}
\begin{table}[b]
\centering
\caption{
Modified NDCG score for $k=5$ and $k=10$.
}
\label{tab:retrieval}
\scriptsize
\begin{tabular*}{\textwidth}{
l
*{3}{@{\extracolsep{\fill} }c}
*{3}{@{\extracolsep{\fill} }c}
*{2}{@{\extracolsep{\fill} }c}
*{2}{@{\extracolsep{\fill} }c}
}
\toprule
& 
\multicolumn{3}{c}{\cite{dumani2020framework}} &
\multicolumn{3}{c}{top-k} &
\multicolumn{2}{c}{k-Means} &
\multicolumn{2}{c}{\acrshort{biased-coreset}} 
\\
k & first & sent & sliding & zero-shot & same topic & ours & BERT & CLAIM-SIM & BERT & CLAIM-SIM \\
\midrule
5 & .399 & .378  & .455 & .437 & .373 & .447 & .428 & .465 & .437 & \textbf{.475} \\
10 & .455 & .429 & .487 & .476 & .448 &.502 & .515 & .513 & .520 & \textbf{.526}\\
\end{tabular*}

\end{table}
\paragraph{Experimental Setting} The training dataset of the relevance filter is a subset of 160,000 positive (relevant) claim-premise sentence pairs of the dataset described in~\cite{dumani2019systematic}.
Additionally, we generated 320,000 negatives (not-relevant) claim-premise pairs as described in Section~\ref{relevance_filter}.
For the evaluation of our approach and comparison with the baselines, we utilize the dataset from~\cite{dumani2020framework}. 
The evaluation set consists of 1,195 triples ($c_{query}, c_{result}, p_{result}$) each labeled as "very relevant" (389), "relevant" (139) or "not relevant" (667). 
The 528 "very relevant" and "relevant" premises were assigned to groups with the same meaning by human annotators.
In contrast to \cite{dumani2020framework} we do \emph{not} utilize the ground truth assignments of $c_{result} \leftrightarrow p_{result}$ in our approach. Therefore our method can utilize newly arriving premises without an assignment to $c_{result}$.
To select the optimal hyperparameters for our approach and avoid test leakage, we use leave-one-out cross-validation:
For each query claim with corresponding premises, we use the rest of the evaluation dataset to select the hyperparameters and then evaluate this hold-out query.
To obtain a final score, we average over all splits.
As an evaluation metric, we use the modified nDCG from~\cite{dumani2020framework}:
Only the first occurrence from a premise ground truth cluster yields positive gain; duplicates do not give any gain.
In Table~\ref{tab:retrieval}, we summarize the results of the argument retrieval task.
The numbers represent the modified NDCG scores for $k=5$ and $k=10$.
The first three columns show the evaluation results for the methods from~\cite{dumani2020framework}.\footnote{
For the evaluation, we have used interim results provided by the authors of the original publication. 
Since we had obtained deviations from the originally reported results, we have contacted the authors and came together to the conclusion that our numbers are correct. 
We thank the authors for their help.
}
In the next three columns denoted as \emph{top-k}, we present the results when premises with the highest score are returned directly, without de-duplication. 
With the \emph{zero-shot} approach, we investigate the assumption that similarity between query and claim is not a sufficient indicator for relevance.
Thus, we use the similarity between representations obtained from a pre-trained BERT model without training on claim-premise relevance.
The second column, \emph{same topic}, denotes the performance of the relevance model trained in the same setting as our approach with the only difference that negative instances for the training are selected from the same topic.
Finally, \emph{ours} denotes the setting, where $k$ instances have the highest probability to be relevant estimated by our model (more precisely, the \emph{relevance filter}). 
Given these results, we observe a strong performance of the \emph{zero-shot} approach, which comes close to the approaches by \cite{dumani2020framework}. 
We emphasize that this is even though this baseline approach neither uses ground truth premise-claim relevance data as \cite{dumani2020framework}, nor any other external premise-claim relevance data.
Moreover, we observe that we can achieve good performance in terms of the \emph{modified} NDCG despite not filtering duplicates.
At the same time, we observe that our model can still improve the similarity-based approach by several points.
In contrast, the model learned with negatives instances from the same topic performs much worse than \emph{zero-shot}, which underlines the correct task's importance.
Finally, the columns denoted as \emph{Biased Coreset} present our final results.
The results are from the \emph{premise ranker} applied to the different premise representations of the most relevant premises selected by \emph{relevance filter}.
For comparison, we also report the results, where k-means is used as \emph{premise ranker} on the same representations, where we select at most one premise per cluster according to the similarity.
The \emph{claim-sim} premise representation always outperforms \emph{bert} and our \emph{biased-coreset} premise ranker is better than the k-means clustering.

\section{Conclusion}
In this work, we have presented a novel approach for the retrieval of \emph{relevant} and \emph{original} premises for the query claims. Our new approach can be applied more flexibly than previous methods since it does not require mappings between premises and claims in the database. 
Thus, it can also be applied in an inductive setting, where new premises can be used without the need first to associate them with relevant claims manually.
At the same time, it achieves better results than approaches that make use of this information.

\section{Acknowledgments}
This work has been funded by the German Federal Ministry of Education and Research (BMBF) under Grant No. 01IS18036A and by the Deutsche Forschungsgemeinschaft (DFG) within the project Relational Machine Learning for  Argument  Validation  (ReMLAV),  Grant  NumberSE  1039/10-1,  as part of the Priority  Program  "Robust Argumentation Machines (RATIO)" (SPP-1999). The authors of this work take full responsibility for its content.

\bibliographystyle{splncs04}
\bibliography{main}

\begin{thebibliography}{10}
\providecommand{\url}[1]{\texttt{#1}}
\providecommand{\urlprefix}{URL }
\providecommand{\doi}[1]{https://doi.org/#1}

\bibitem{akiki:2020}
Akiki, C., Potthast, M.: {Exploring Argument Retrieval with Transformers}. In:
  Working Notes Papers of the CLEF 2020 Evaluation Labs (Sep 2020)

\bibitem{amati2002probabilistic}
Amati, G., Van~Rijsbergen, C.J.: Probabilistic models of information retrieval
  based on measuring the divergence from randomness. ACM Transactions on
  Information Systems (TOIS)  \textbf{20}(4),  357--389 (2002)

\bibitem{wachsmuth2020overview}
Bondarenko, A., Fr{\"o}be, M., Beloucif, M., Gienapp, L., Ajjour, Y.,
  Panchenko, A., Biemann, C., Stein, B., Wachsmuth, H., Potthast, M., Hagen,
  M.: {Overview of Touch{\'e} 2020: Argument Retrieval}. In: Cappellato, L.,
  Eickhoff, C., Ferro, N., N{\'e}v{\'e}ol, A. (eds.) Working Notes Papers of
  the CLEF 2020 Evaluation Labs. CEUR Workshop Proceedings, vol.~2696 (Sep
  2020), \url{http://ceur-ws.org/Vol-2696/}

\bibitem{chernodub2019targer}
Chernodub, A., Oliynyk, O., Heidenreich, P., Bondarenko, A., Hagen, M.,
  Biemann, C., Panchenko, A.: Targer: Neural argument mining at your
  fingertips. In: Proceedings of the 57th Annual Meeting of the Association for
  Computational Linguistics: System Demonstrations. pp. 195--200 (2019)

\bibitem{devlin-etal-2019-bert}
Devlin, J., Chang, M.W., Lee, K., Toutanova, K.: {BERT}: Pre-training of deep
  bidirectional transformers for language understanding. In: Proceedings of the
  2019 Conference of the North {A}merican Chapter of the Association for
  Computational Linguistics: Human Language Technologies, Volume 1 (Long and
  Short Papers). pp. 4171--4186. Association for Computational Linguistics,
  Minneapolis, Minnesota (Jun 2019). \doi{10.18653/v1/N19-1423},
  \url{https://www.aclweb.org/anthology/N19-1423}

\bibitem{dumani2020framework}
Dumani, L., Neumann, P.J., Schenkel, R.: A framework for argument retrieval.
  In: European Conference on Information Retrieval. pp. 431--445. Springer
  (2020)

\bibitem{dumani2019systematic}
Dumani, L., Schenkel, R.: A systematic comparison of methods for finding good
  premises for claims  (2019)

\bibitem{ein2020corpus}
Ein-Dor, L., Shnarch, E., Dankin, L., Halfon, A., Sznajder, B., Gera, A.,
  Alzate, C., Gleize, M., Choshen, L., Hou, Y., et~al.: Corpus wide argument
  mining-a working solution. In: AAAI. pp. 7683--7691 (2020)

\bibitem{feger2020structure}
Feger, M., Steimann, J., Meter, C.: Structure or content? towards assessing
  argument relevance. In: Proceedings of the 8th International Conference on
  Computational Models of Argument (COMMA 2020). p.~135 (2020)

\bibitem{DBLP:conf/webi/FrommF019}
Fromm, M., Faerman, E., Seidl, T.: {TACAM:} topic and context aware argument
  mining. In: Barnaghi, P.M., Gottlob, G., Manolopoulos, Y., Tzouramanis, T.,
  Vakali, A. (eds.) 2019 {IEEE/WIC/ACM} International Conference on Web
  Intelligence, {WI} 2019, Thessaloniki, Greece, October 14-17, 2019. pp.
  99--106. {ACM} (2019). \doi{10.1145/3350546.3352506},
  \url{https://doi.org/10.1145/3350546.3352506}

\bibitem{habernal-gurevych-2016-argument}
Habernal, I., Gurevych, I.: Which argument is more convincing? analyzing and
  predicting convincingness of web arguments using bidirectional {LSTM}. In:
  Proceedings of the 54th Annual Meeting of the Association for Computational
  Linguistics (Volume 1: Long Papers). pp. 1589--1599. Association for
  Computational Linguistics, Berlin, Germany (Aug 2016).
  \doi{10.18653/v1/P16-1150}, \url{https://www.aclweb.org/anthology/P16-1150}

\bibitem{kriegel2009clustering}
Kriegel, H.P., Kr{\"o}ger, P., Zimek, A.: Clustering high-dimensional data: A
  survey on subspace clustering, pattern-based clustering, and correlation
  clustering. ACM Transactions on Knowledge Discovery from Data (TKDD)
  \textbf{3}(1),  1--58 (2009)

\bibitem{reimers-etal-2019-classification}
Reimers, N., Schiller, B., Beck, T., Daxenberger, J., Stab, C., Gurevych, I.:
  Classification and clustering of arguments with contextualized word
  embeddings. In: Proceedings of the 57th Annual Meeting of the Association for
  Computational Linguistics. pp. 567--578. Association for Computational
  Linguistics, Florence, Italy (Jul 2019). \doi{10.18653/v1/P19-1054},
  \url{https://www.aclweb.org/anthology/P19-1054}

\bibitem{sener2017active}
Sener, O., Savarese, S.: Active learning for convolutional neural networks: A
  core-set approach. arXiv preprint arXiv:1708.00489  (2017)

\bibitem{stab-etal-2018-cross}
Stab, C., Miller, T., Schiller, B., Rai, P., Gurevych, I.: Cross-topic argument
  mining from heterogeneous sources. In: Proceedings of the 2018 Conference on
  Empirical Methods in Natural Language Processing. pp. 3664--3674. Association
  for Computational Linguistics, Brussels, Belgium (Oct-Nov 2018).
  \doi{10.18653/v1/D18-1402}, \url{https://www.aclweb.org/anthology/D18-1402}

\bibitem{Staudte2020SentArgAH}
Staudte, C., Lange, L.: Sentarg: A hybrid doc2vec/dph model with sentiment
  analysis refinement. In: CLEF (2020)

\bibitem{trautmann2020relational}
Trautmann, D., Fromm, M., Tresp, V., Seidl, T., Sch{\"u}tze, H.: Relational and
  fine-grained argument mining. Datenbank-Spektrum pp.~1--7 (2020)

\bibitem{vaswani2017attention}
Vaswani, A., Shazeer, N., Parmar, N., Uszkoreit, J., Jones, L., Gomez, A.N.,
  Kaiser, {\L}., Polosukhin, I.: Attention is all you need. In: Advances in
  neural information processing systems. pp. 5998--6008 (2017)

\bibitem{wachsmuth2017computational}
Wachsmuth, H., Naderi, N., Hou, Y., Bilu, Y., Prabhakaran, V., Thijm, T.A.,
  Hirst, G., Stein, B.: Computational argumentation quality assessment in
  natural language. In: Proceedings of the 15th Conference of the European
  Chapter of the Association for Computational Linguistics: Volume 1, Long
  Papers. pp. 176--187 (2017)

\bibitem{wachsmuth2017building}
Wachsmuth, H., Potthast, M., Al~Khatib, K., Ajjour, Y., Puschmann, J., Qu, J.,
  Dorsch, J., Morari, V., Bevendorff, J., Stein, B.: Building an argument
  search engine for the web. In: Proceedings of the 4th Workshop on Argument
  Mining. pp. 49--59 (2017)

\end{thebibliography}
\end{document}